\documentclass[aps,prl,reprint,showpacs,superscriptaddress]{revtex4-1}

\usepackage{graphicx}
\usepackage{dcolumn}
\usepackage{bm}
\usepackage{amsmath,amssymb}
\usepackage{comment}

\def\U#1{{%
\def\O{\mbox{O}}
\def\u{\mbox{u}}
\mathcode`\u=\mu
\mathcode`\O=\Omega
\mathrm{#1}}}

\def\Im{\mathop{\mathrm{Im}}}

\def\ii{{\mathrm{i}}}

\def\sub#1{_{\scriptsize\mbox{#1}}}

\begin{document}


\title{Suppression of narrow-band transparency in a metasurface induced\\
by a strongly enhanced electric field}

\author{Yasuhiro Tamayama}
\email{tamayama@vos.nagaokaut.ac.jp}
\author{Keisuke Hamada}
\author{Kanji Yasui}
\affiliation{Department of Electrical, Electronics and
Information Engineering, Nagaoka University of
Technology, 1603-1 Kamitomioka, Nagaoka, Niigata 940-2188, Japan}

\date{\today}

\begin{abstract}

We realize a suppression of an electromagnetically induced
transparency (EIT) like transmission in a metasurface induced by
a local electric field that is strongly enhanced based on two approaches: 
squeezing of electromagnetic energy in resonant metasurfaces and 
enhancement of electromagnetic energy density associated with a low group
velocity.
The EIT-like metasurface consists of a pair of radiatively coupled
cut-wire resonators, and it can effect both field enhancement 
approaches simultaneously. 
The strongly enhanced local electric field generates an air discharge
plasma at either of the gaps of the cut-wire resonators, which causes 
the EIT-like metasurface to change into two kinds of Lorentz type metasurfaces. 

\end{abstract}

\maketitle

\section{I. Introduction}

There has been considerable interest in controlling electromagnetic
waves using metamaterials. Metamaterials are artificial continuous media
made of arrays of subwavelength structures. The electromagnetic response
of a metamaterial is determined by the shape, material, and distribution
of its unit structure, and therefore metamaterials with desired
characteristics can be realized by designing the structure of the unit cell. 

One of the most promising applications of metamaterials is the realization of
highly nonlinear media~\cite{lapine14}. 
In a metamaterial composed of resonant structures such as split-ring
resonators, electromagnetic energy is squeezed into small volume regions 
of the metamaterial~\cite{pendry99}.
Nonlinear elements located in these regions exhibit enhanced nonlinear
responses owing to the compression of the electromagnetic field.
To date, enhancement of harmonic 
generations~\cite{klein06,shadrivov08_apl,kim08,rose11,kanazawa11,nakanishi12,czaplicki13,obrien15}, 
bistable media~\cite{wang_b08,lapine12,liu_m_prb13},
tunable media~\cite{shadrivov08_apl,wang_b08,powell09,poutrina10}, and
modulation instability~\cite{tamayama13}
have been experimentally investigated using resonant metamaterials. 

Another strategy for enhancing nonlinear phenomena is to decrease 
the energy velocity, which is equal to the group velocity of electromagnetic waves in lossless
media~\cite{krauss08}. 
The electromagnetic energy density increases with decreasing
energy velocity~\cite{chen_pra08} and therefore the interaction between the
electromagnetic wave and the nonlinear medium is enhanced.

If the above two approaches for enhancing nonlinear phenomena are
integrated, nonlinear phenomena can be further enhanced. 
Although the effects of the two approaches are not clearly separable in resonant
metamaterials, the resonant enhancement of nonlinearity can be maximized
by using the group velocity as a index of the field enhancement. 
In this study, as a verification of this concept, 
we realize a change of an electromagnetic response
of a metasurface induced
by a strongly enhanced electromagnetic field. 
A metasurface that mimics electromagnetically induced transparency
(EIT)~\cite{fedotov07,zhang_prl08,tassin_prl09,liu_nat09,tamayama10,zhang_apl_10,kurter11,tamayama12,gu12,nakanishi13,miyamaru14,tamayama14}
is used to simultaneously realize squeezing 
of the electromagnetic energy and a low group velocity.
The response of the EIT-like metasurface is theoretically analyzed
based on an electrical circuit model of its unit structure to show the
condition for minimizing the group velocity, i.e., maximizing the
electric field enhancement factor.
The maximum enhancement factor is examined through
measurement and numerical analysis of the linear characteristics of
the metasurface. 
The strongly enhanced electric field generates an air
discharge plasma in the metasurface, which causes a change in its
effective structure.

\section{II. Theory}

\begin{figure}[tb]
\begin{center}
\includegraphics[scale=0.7]{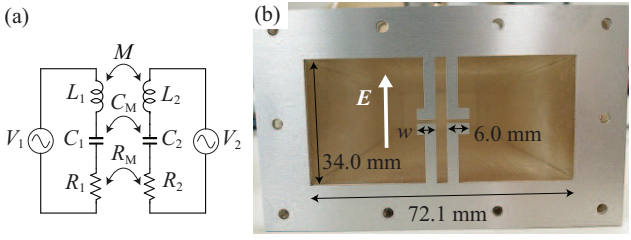}
\caption{(a) Electrical circuit model for the unit cell of the EIT-like
 metasurface. 
 (b) Photograph of the fabricated metasurface. }
\label{fig:structure}
\end{center}
\end{figure}

The characteristics of the EIT-like metasurface
used in this study are theoretically 
analyzed based on the electrical circuit model of its unit structure
shown in Fig.\,\ref{fig:structure}(a) to clarify the method for
decreasing the group velocity, i.e., enhancing the local electric
field. The electrical
circuit consists of two inductor-capacitor series resonant circuits 
coupled via a mutual impedance 
$Z\sub{M} = R\sub{M} - \ii [\omega M - (\omega C\sub{M})^{-1}]$, where
$\omega$ is the angular frequency of the voltage sources that correspond
to the incident wave.
The following assumptions are made in the analysis below to simplify the
calculation. 
(1) The circuit constants satisfy $L_1 = L_2 = L_0$, $C_1 \simeq C_2$,
$R_1 = R_2 =R_0$, $V_1 = V_2$, and $\Im{(Z\sub{M})}=0$. 
The first four conditions imply that the characteristics of the two
resonators are the same except for the resonant frequency and that the
resonant frequencies are close to each other. 
The last equation implies that the resonators
are coupled only via indirect coupling, i.e., radiative
coupling~\cite{zhang_s_12,verslegers12}.
(2) The electric dipole moment of the unit cell
of the metasurface is
proportional to the sum of the stored charges in the two capacitors.
Applying Kirchhoff's voltage law to the electrical circuit and using the
above assumptions, 
the electric susceptibility $\chi\sub{e}$ of the metasurface 
at around $\omega \simeq \omega _0$ can be written
as follows:
\begin{equation}
\chi\sub{e}
\approx
\frac{-\alpha ( \omega^2 - \omega_0^2 + \ii \gamma\sub{L} \omega)}{
(\omega^2 - \omega_0^2 + \ii \gamma_0 \omega)^2 
- [(\omega_2^2 - \omega_1^2)/2]^2 + (\omega R\sub{M} / L_0 )^2 }
, \label{eq:40}
\end{equation}
where $\omega_{1,2} = 1/\sqrt{L_0 C_{1,2}}$, 
$\omega_0^2 = (\omega_1^2 + \omega_2^2 ) / 2$, 
$\gamma_{0} = R_0 / L_0$, 
$\gamma\sub{L} = \gamma_0 - (R\sub{M}/L_0)$, and 
$\alpha$ is a proportionality constant. 
The right-hand side of Eq.\,(\ref{eq:40}) is similar to the
electric susceptibility of EIT~\cite{tamayama14}. The EIT-like 
transparency occurs at an incident
angular frequency of $\omega_0 \approx (\omega_1 + \omega_2)/2$. 
Note that $\gamma\sub{L}$ represents the dissipated power from the
resonators that does not contribute to the indirect coupling, because
$\gamma_0$ represents the total power dissipated from each resonator
and $R\sub{M}/L_0$ represents the radiative coupling between the
resonators. 
Assuming that the group index at $\omega = \omega_0$ is much larger than
the refractive index, 
the group index at $\omega = \omega_0$ is then written as follows: 
\begin{equation}
n\sub{g} \approx 
\frac{\alpha (\varDelta^2 - \gamma\sub{L}^2)}{(\varDelta^2 + 2\gamma_0
\gamma\sub{L} - \gamma\sub{L}^2)^2}, \label{eq:260} 
\end{equation}
where $\varDelta = | \omega_1 - \omega_2 |$.
When $\gamma\sub{L} \ll \gamma_0$, 
the right-hand side of Eq.\,(\ref{eq:260}) exhibits a maximum value of
$\alpha / 8 \gamma_0 \gamma\sub{L}$ at
$\varDelta = \sqrt{2\gamma_0 \gamma\sub{L}} = \varDelta\sub{max}$. 
It can be confirmed from a simple calculation 
that the stored charge in each capacitor at the transparency
frequency also exhibits a maximum 
at $\varDelta = \varDelta\sub{max}$. This implies that the local
electric field in the metasurface becomes largest when the group
index is maximized. 

It is found from the above analysis that a reduction of
$\gamma\sub{L}$ is essential for increasing the maximum 
group index and the electric field enhancement factor. 
In order to decrease $\gamma\sub{L}$, 
nonradiative losses such as dielectric loss and Ohmic loss should be
suppressed and the radiation modes of the two resonators should be
similar to each other. 

We designed and fabricated the EIT-like metasurface shown in
Fig.\,\ref{fig:structure}(b) based on the above considerations. 
The structure was cut from an aluminum plate with a thickness of
$1.0\,\U{mm}$ using wire electrical discharge machining. 
The unit cell of the metasurface 
consists of two kinds of cut-wire resonators. 
The resonator at the right-hand side is referred to as
resonator L and the left-hand side resonator is referred to as resonator H. 
(In this experiment, the dimension $w$ in resonator H, indicated in
Fig.\,\ref{fig:structure}(b), is smaller than $6.0\,\U{mm}$, the length
of the corresponding dimension in resonator L, 
and the resonant frequency of
resonator L is therefore lower than that of resonator H.)  
The two cut-wire resonators are arranged so that 
the radiation from one resonator can excite the other resonator, that
is, the two resonators are indirectly coupled.
Since the resonators resembles each other in shape,
their radiation modes are similar. 
The metasurface is made only of aluminum, which behaves
as a perfect electric conductor in the microwave region, and thus 
no dielectric or Ohmic loss occur in the metasurface.
Therefore, a small $\gamma\sub{L}$ can be achieved in this metasurface. 

\section{III. Linear characteristics of the metasurface}

\begin{figure*}[tb]
\begin{center}
\includegraphics[scale=0.7]{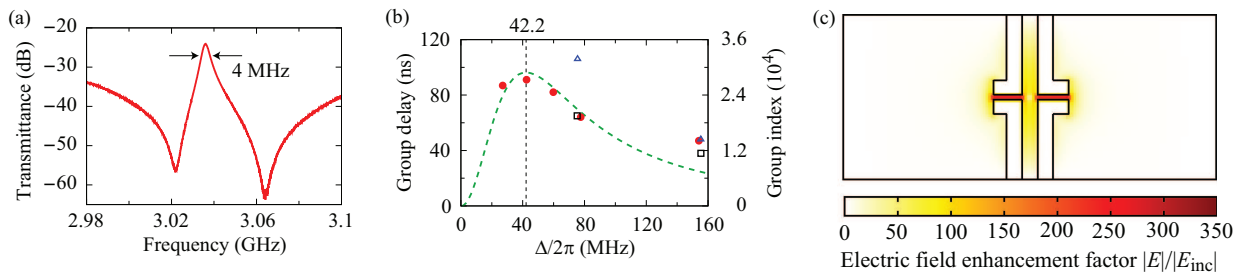}
\caption{(a) Linear transmission spectrum for $w=5.7\,\U{mm}$. (b)
 Measured group delay at the transparency frequency for
 $w=5.0\,\U{mm}$, $5.5\,\U{mm}$, $5.6\,\U{mm}$, $5.7\,\U{mm}$,
 and $5.8\,\U{mm}$ in order from right to left 
 (solid circles) and theoretical fit according to
 Eq.\,(\ref{eq:260}) (dashed curve). 
 The open squares and triangles represent the numerically analyzed values for 
 $\sigma = 5.0 \times 10^4 \,\U{S/m}$ 
 and $1.0 \times 10^5 \,\U{S/m}$, respectively. 
 (c) Distribution of numerically calculated ratio of the electric field
 amplitude $|E|$ to the incident electric field amplitude $|E\sub{inc}|$ for
 $w=5.7\,\U{mm}$ and $\sigma = 5.0\times 10^4\,\U{S/m}$.}
\label{fig:linear}
\end{center}
\end{figure*}

First, we measured the linear transmission characteristics of the fabricated
metasurfaces with various $w$
to confirm the above theory and to find the condition 
that realizes the maximum group index, i.e., the maximum local
electric field enhancement. The metasurface was placed in a rectangular
waveguide and the transmission and group delay spectra of the metasurface 
were measured using a network
analyzer. A thru-reflection-load (TRL) calibration~\cite{trl_cal} was used to
eliminate errors caused by multiple reflections and transmission losses 
in the experimental system. 

Figure \ref{fig:linear}(a) shows the measured transmission spectrum of
the metasurface with $w=5.7\,\U{mm}$. 
An EIT-like transmission window with a center frequency of
$3.036\,\U{GHz}$ and a bandwidth of $4\,\U{MHz}$ is observed. 
The difference between the transmission dip frequencies, which are
almost the same as the resonant frequencies of the two cut-wire
resonators~\cite{tamayama14}, is $42.5\,\U{MHz}$. 
This value is much smaller than the resonance linewidth of the cut-wire
resonators, $\gamma_0 / 2\pi = 920\,\U{MHz}$, which was determined from the decay
rate of the stored energy in the cut-wire resonator calculated using a
finite-difference time-domain method (not shown). 
Thus, $\gamma\sub{L}$ is confirmed to be much smaller than $\gamma_0$. 

Figure \ref{fig:linear}(b) shows the measured 
group delay at the transparency
frequency for five different metasurfaces with
$w=5.0\,\U{mm}$, $5.5\,\U{mm}$, $5.6\,\U{mm}$, $5.7\,\U{mm}$, 
and $5.8\,\U{mm}$ as solid circles in order from right to left. 
Here $\varDelta / 2\pi$ is approximated by the difference between the two
transmission dip frequencies. The measured group delay reaches a maximum
value of $91\,\U{ns}$ at $w=5.7\,\U{mm}$. Assuming that the effective
thickness of the metasurface is equal to the thickness of the aluminum
plate, the maximum group delay corresponds to a group index of 
$2.7 \times 10^4$. 
A theoretical fit of the measured data to Eq.\,(\ref{eq:260}) is shown
by the dashed curve. 
The measured values and fitted curve are in good agreement except for
the case of $w=5.0\,\U{mm}$. This difference may be due to
the assumptions used in deriving Eq.\,(\ref{eq:260}) and 
by the fact that $\gamma\sub{L}$ 
varies among the fabricated samples (as described below). 
The fitted curve exhibits a maximum at 
$\varDelta / 2\pi = 42.2\,\U{MHz}$,
which is almost equal to $\varDelta / 2\pi$ for the case of
$w=5.7\,\U{mm}$. Therefore, the metasurface with 
$w=5.7\,\U{mm}$ is confirmed to have 
the optimum condition for maximizing the group
index and the electric field enhancement factor at the transparency
frequency. 

The electric field distribution in the metasurface was analyzed using a
finite-difference time-domain method
to estimate the electric field enhancement factor. 
The effective conductivity $\sigma$ of the aluminum in this experiment needs to be
evaluated for the calculation of the enhancement factor. 
The calculated group delays at the transparency frequency for 
$\sigma = 5.0 \times 10^4 \,\U{S/m}$ and 
$1.0 \times 10^5 \,\U{S/m}$ are 
shown as the open squares and open triangles, respectively,
in Fig.\,\ref{fig:linear}(b). 
(The calculations were performed only for the cases of
$w=5.0\,\U{mm}$ and $5.5\,\U{mm}$ due to the limitations of memory and
computing time.)
The measured values seem to be in the range of the calculated values for
$\sigma = 5.0 \times 10^4 \,\U{S/m}$ and $1.0 \times 10^5 \,\U{S/m}$. 
Thus, the effective conductivity of the aluminum in this experiment is of the
order of $5 \times 10^4 \,\U{S/m}$ to $1 \times 10^5 \,\U{S/m}$. 
Figure \ref{fig:linear}(c) shows the numerically calculated electric
field enhancement factor for $w=5.5\,\U{mm}$ and 
$\sigma = 5.0 \times 10^4 \,\U{S/m}$. The maximum enhancement factor of
about $300$ is observed at the gaps of the cut-wire resonators. 
The enhancement factor for $w=5.7\,\U{mm}$ would be larger than this value
because 
the group delay for $w=5.7\,\U{mm}$ is larger than that for
$w=5.5\,\U{mm}$. 

Here we discuss the meaning of the effective conductivity in the above
simulation. The effective conductivity represents the leak of the
indirect coupling as well as the Ohmic loss. If the surface roughness of the
cut edge of the metasurface is not much smaller than the skin depth of
the metal, the surface current flow is affected by the surface
roughness. As a result, mismatch between the radiation modes of the two
cut-wire resonators increases and the leak of the indirect coupling
increases. That is, $\gamma\sub{L}$ in Eqs.\,(\ref{eq:40}) and
(\ref{eq:260}) increases. It is difficult to
directly deal with the surface roughness of the metal in the simulation,
while both of the leak of the indirect coupling and the Ohmic loss are the
factors that increase $\gamma\sub{L}$. Therefore, the leak caused
by the surface roughness is treated as a factor that decreases the
conductivity in the simulation. Since the surface roughness may vary
among the fabricated samples, the effective conductivity varies among
the samples. In fact, we have already confirmed that $\gamma\sub{L}$ varies
between the samples with the same geometrical parameters that were
fabricated using different methods. The effective conductivity in this
experiment is much smaller than the conductivity of bulk aluminum,
$3.8 \times 10^7 \,\U{S/m}$.
This implies that the leak of the indirect coupling caused by
the surface roughness is much larger than the Ohmic loss.

\section{IV. Nonlinear characteristics of the metasurface}

\begin{figure}[tb]
\begin{center}
\includegraphics[scale=0.5]{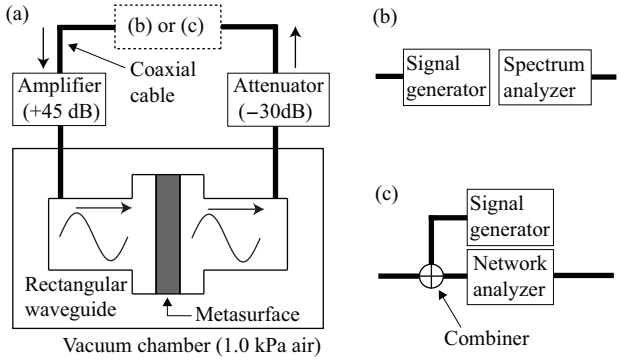}
\caption{(a) Experimental setup for the measurement of the nonlinear
 characteristics of the metasurface. 
 The dashed rectangle represents the setup shown in (b) and (c). 
 (b) Setup for the continuous wave measurement and 
 (c) setup for the pump-probe measurement. }
\label{fig:setup}
\end{center}
\end{figure}

Next, we investigated the nonlinear response of the metasurface 
induced by the strongly enhanced electric field. 
Hereafter, the metasurface with $w=5.7\,\U{mm}$ is used. 
The experimental setup is shown in Figs.\,\ref{fig:setup}(a) and
\ref{fig:setup}(b). 
A continuous wave generated from a microwave signal generator
was amplified and then fed into the rectangular waveguide. 
The wave was incident onto the metasurface in the waveguide. 
The transmitted wave was attenuated and then fed into a spectrum
analyzer. The rectangular waveguide was placed in a vacuum chamber made
of acrylic where
the pressure of air was reduced to $1.0 \,\U{kPa}$. 
The TRL calibration was not carried out in this experiment 
and the total 
transmittance from the output port of the signal generator to the input
port of the spectrum analyzer was measured. 

\begin{figure}[tb]
\begin{center}
\includegraphics[scale=0.7]{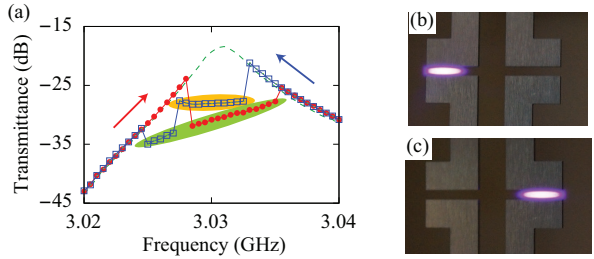}
\caption{(a) Transmission spectra of the metasurface with 
 $w=5.7\,\U{mm}$ for an incident power of $7.9\,\U{W}$. The red solid
 circles and blue open squares 
 represent the transmittance when the incident frequency is swept from
 low to high and from high to low frequencies, respectively. 
 When transmittance in the yellow or green region is observed, the
 discharge occurs in resonator H or resonator L, respectively.
 The solid lines are guides for the eye.
 The green dashed curve represents the linear transmission spectrum. 
 (b) and (c) Photographs of the metasurface when the discharge occurs
 in resonator H and resonator L, respectively.}
\label{fig:nonlinear}
\end{center}
\end{figure}

Figure \ref{fig:nonlinear}(a) shows the measured 
transmission spectra of the metasurface for an incident power of
$7.9\,\U{W}$.
The red solid circles and blue open squares 
represent the transmittance
when the incident frequency is swept from low to high and from high to low
frequencies, respectively. 
(Harmonic generations were not observed in this experiment.)
The green dashed curve represents the linear transmission
spectrum that was measured for an incident power below $2.8\,\U{W}$, 
which corresponds to an incident
electric field amplitude of $15\,\U{V/cm}$. 
There is a frequency region where
the transmittance becomes smaller than the linear
transmittance at around $3.03\,\U{GHz}$. 
This observation implies that discharge of air occurs 
in the metasurface in this frequency region. 
One of the gaps of 
the cut-wire resonators is filled by an air plasma, which behaves as a 
conductor, because of the discharge; thus, the transparency condition
collapses and the transmittance decreases. 
When transmittance in the yellow or green region is observed, the discharge occurs in
resonator H or resonator L, respectively. 
This is because the electric field enhancement is larger in resonator H
than in resonator L for a frequency
larger
than the transmission peak frequency, and vice versa for a frequency smaller than the transmission peak frequency. 
The discharge starts in the resonator if the local electric field
enhancement in the resonator 
is larger than that in the other resonator and the local
electric field exceeds the breakdown electric field. 
Figures \ref{fig:nonlinear}(b) and \ref{fig:nonlinear}(c) show 
photographs of the 
metasurface when the discharge occurs in resonator H and resonator L, 
respectively. 
Purple light emission due to the discharge is observed. 
The results of this experiment indicate that the 
discharge can be induced at either of
the gaps by controlling the frequency
and power of the incident wave owing to the enhancement
of the local electric field.

Finally, the transmission
spectrum of the metasurface for a probe wave was measured under
irradiation by a pump wave that controls the discharge in the
metasurface. 
The experimental setup is shown in Figs.\,\ref{fig:setup}(a) and
\ref{fig:setup}(c).
A signal generator was used to generate the pump wave and 
a network analyzer was used to generate and detect the
probe wave. The power of the probe wave was set to be small enough not
to affect the discharge state. 

\begin{figure}[tb]
\begin{center}
\includegraphics[scale=0.7]{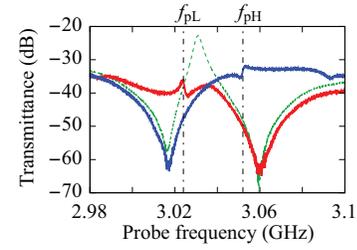}
\caption{Transmission spectra of the metasurface with $w=5.7\,\U{mm}$ 
 for the probe wave 
 when the discharge is induced in resonator L (red solid curve) and resonator H (blue solid curve) 
 by the pump wave with frequency of $f\sub{pL} = 3.024\,\U{GHz}$ and
 $f\sub{pH} = 3.052\,\U{GHz}$, respectively. The pump power is $15.8\,\U{W}$.
 The green dashed curve represents the transmission spectrum without the
 pump wave. }
\label{fig:pump}
\end{center}
\end{figure}

Figure \ref{fig:pump} shows examples of the measured 
transmission spectra of the probe wave for a pump power of
$15.8\,\U{W}$. 
The red (blue) solid curve shows the spectrum when the discharge occurs
in resonator L (resonator H) and the pump frequency is $3.024\,\U{GHz}$
($3.052\,\U{GHz}$). 
The green dashed curve is the spectrum without the pump wave. 
Note that the transmittance in this figure represents the total 
transmittance from the transmitting port to the receiving
port of the network analyzer. 
When a discharge occurs in the metasurface, one of the transmission
dips disappears, which causes the transmission peak to disappear. 
One of the cut-wire resonators changes to resemble a continuous wire, which
is a non-resonant structure, because of the
air plasma created by the
discharge. 
Only the resonant spectrum of the cut-wire resonator where the
discharge does not occur is observed. 
That is, the EIT-like
metasurface is changed into two kinds of Lorentz-type metasurfaces by
controlling the frequency and power of the pump wave. 
In this experiment, simultaneous disappearance of both transmission dips
is not observed at any pump frequencies even for a pump power of
$15.8\,\U{W}$, which is $5.6$ times as large as the threshold incident power for the
discharge. This implies that the discharge does not simultaneously occur
at both gaps of the cut-wire resonators for an incident power of
$15.8\,\U{W}$. It is confirmed from this observation that the electric field
enhancement in the metasurface is stronger than that in each constituent
resonator, which is caused by the integration of the two approaches for
the field enhancement.

The above results show that the metasurface exhibits the
EIT-like (Lorentz type) response when the pump wave is turned off
(on). The transmission of the probe wave can be
controlled by the pump wave like in the original EIT though the
dependence of the transmission on the pump wave is opposite to the
EIT. So to speak, electromagnetically induced suppression of narrowband
transparency is realized in the metasurface. It may be possible to store
the probe wave in the metasurface by controlling the pump wave as in the
case of the EIT.

Note that all the experiments in this study are highly
reproducible. Any deterioration of the characteristics of the
metasurface has not been caused by the discharge in our experiment where
the incident power is less than $15.8\,\U{W}$ (which is the maximum
output power of the used amplifier). 

\section{V. Conclusion}

In conclusion,
we have realized the suppression of an EIT-like transmission
in the metasurface
induced by a strongly enhanced local electric field.
To strongly enhance the local
electric field, we integrated two approaches: 
squeezing of electromagnetic energy in
resonant metamaterials and enhancement of electromagnetic energy
density associated with a low group velocity. 
The EIT-like 
metasurface composed of a pair of indirectly coupled cut-wire resonators was 
used as a metasurface for which the above two approaches can be used
simultaneously.
The structure of the metasurface was designed and fabricated based on
the theoretical analysis of the electrical circuit model of the
EIT-like metasurface. 
An electric field enhancement factor of about
$300$ was achieved by maximizing the group delay in the
metasurface. 
Owing to the strongly enhanced local electric field, discharge could occur
at either of the gaps of the cut-wire resonators. 
A pump-probe experiment demonstrated that 
the EIT-like metasurface was changed
into two kinds of Lorentz-type metasurfaces by controlling the power
and frequency of the pump wave.
The present metasurface can be applied to limiters and switches for
electromagnetic waves. 
It may also be possible to realize memories for electromagnetic waves
and to efficiently generate various nonlinear phenomena. 
In addition, this metasurface could be used for the simple 
generation of plasma, which can contribute to material processing,
chemical reactions, and other applications. 


\acknowledgments

This research was supported by a Grant-in-Aid for Scientific Research on
Innovative Areas (No.\@ 22109004) from the Ministry of Education,
Culture, Sports, Science, and Technology of Japan, and by a Grant-in-Aid
for Research Activity Start-up (No.\@ 25889028) from the Japan Society
for the Promotion of Science. 


%


\end{document}